\documentclass[graybox]{svmult}

\usepackage{mathptmx}       % selects Times Roman as basic font
\usepackage{helvet}         % selects Helvetica as sans-serif font
\usepackage{courier}        % selects Courier as typewriter font
\usepackage{type1cm}        % activate if the above 3 fonts are
\usepackage{makeidx}         % allows index generation
\usepackage[dvips]{graphicx}        % standard LaTeX graphics tool
\usepackage{multicol}        % used for the two-column index
\usepackage[bottom]{footmisc}% places footnotes at page bottom

\makeindex             % used for the subject index
\begin{document}

\title*{Economic decision making: application of the theory of complex systems}
\author{Robert Kitt}
\institute{Robert Kitt 
\at Department of Mechanics and Applied Mathematics, Institute of Cybernetics
at Tallinn University of Technology, Akadeemia tee 21, 12061, Tallinn, ESTONIA, \email{kitt@ioc.ee}
\at Swedbank AS, Liivalaia 8, 15040, Tallinn, ESTONIA, \email{robert.kitt@swedbank.ee}}

\maketitle

\abstract*{In this chapter the complex systems are discussed in the context of economic and business policy and decision making. It will be showed and motivated that
social systems are typically chaotic, non-linear and/or non-equilibrium and therefore
complex systems. It is discussed that the rapid
change in global consumer behaviour is underway, that further increases
the complexity in business and management. For policy making
under complexity, following principles are offered: openness and
international competition, tolerance and variety of ideas, self-reliability
and low dependence on external help. 
The chapter contains four applications that build on the theoretical motivation of complexity in social systems. 
The first application demonstrates that small economies have good prospects to gain from the
global processes underway, if they can demonstrate production flexibility,
reliable business ethics and good risk management. 
The second application elaborates on and discusses the opportunities and challenges in decision making under complexity from macro and micro economic perspective. In this environment,
the challenges for corporate management are being also permanently
changed: the balance between short term noise and long term chaos
whose attractor includes customers, shareholders and employees must
be found. The emergence of chaos in economic relationships is demonstrated by a simple system of differential equations that relate the stakeholders described above. 
The chapter concludes with two financial applications:
about debt and risk management. The non-equilibrium economic establishment
leads to additional problems by using excessive borrowing; unexpected
downturns in economy can more easily kill companies. Finally, the
demand for quantitative improvements in risk management is postulated.
Development of the financial markets has triggered non-linearity to
spike in prices of various production articles such as agricultural
and other commodities that has added market risk management to the
business model of many companies.}

\abstract{In this chapter the complex systems are discussed in the context of economic and business policy and decision making. It will be showed and motivated that
social systems are typically chaotic, non-linear and/or non-equilibrium and therefore
complex systems. It is discussed that the rapid
change in global consumer behaviour is underway, that further increases
the complexity in business and management. For policy making
under complexity, following principles are offered: openness and
international competition, tolerance and variety of ideas, self-reliability
and low dependence on external help. 
The chapter contains four applications that build on the theoretical motivation of complexity in social systems. 
The first application demonstrates that small economies have good prospects to gain from the
global processes underway, if they can demonstrate production flexibility,
reliable business ethics and good risk management. 
The second application elaborates on and discusses the opportunities and challenges in decision making under complexity from macro and micro economic perspective. In this environment,
the challenges for corporate management are being also permanently
changed: the balance between short term noise and long term chaos
whose attractor includes customers, shareholders and employees must
be found. The emergence of chaos in economic relationships is demonstrated by a simple system of differential equations that relate the stakeholders described above. 
The chapter concludes with two financial applications:
about debt and risk management. The non-equilibrium economic establishment
leads to additional problems by using excessive borrowing; unexpected
downturns in economy can more easily kill companies. Finally, the
demand for quantitative improvements in risk management is postulated.
Development of the financial markets has triggered non-linearity to
spike in prices of various production articles such as agricultural
and other commodities that has added market risk management to the
business model of many companies.}

%% Robert: abstract with* is used in electronic version; abstract without * is used in book

\section*{Introduction}
\label{sec:intro}
The study of complex systems has long been applied to the social sciences
(cf. \cite{miller,mitchell}). Financial markets and financial time
series have been of special interest among scientists as there is
a well-accessible supply of data in time resolutions ranging from
seconds to years. Therefore, many scientists like physicists, mathematicians,
engineers and others have been attracted by research of the financial
markets. Moreover, for about two decades for now, a branch of statistical
mechanics, \emph{econophysics}, has dealt with the financial time series
analysis (cf. \cite{mantegna,Kitt2005}) by using models first developed
in statistical mechanics (cf. \cite{Stanley} and references therein).

It is not surprising, that the academic literature has been populated with
numerous applications that confirm the non-linearity and/or complexity
in social phenomena. Interestingly, the first two noteworthy applications
came from scholars (as opposed to scientists) in the first half of
20th century: the works of Vilfredo Pareto of wealth distribution
(cf. \cite{pareto}) and George K. Zipf (cf. \cite{zipf}) of frequency
distribution of words in English language. But it was only in 1963, when
Benoit Mandelbrot suggested (cf. \cite{mandel63}) to use Levy stable
distribution function as characteristics of financial market fluctuations;
and in 1965 (cf. \cite{mandel65}) when he recommended processes with long-term
memory. In addition to stochastic phenomena, the deterministic chaos
also describes complexity. One of the first applications of chaos
in economics were reported by T\~onu Puu in 1989 (cf. \cite{Puu}).
The applications of econophysics are mostly descriptive: typically
the dynamical stochastic models are derived from data analysis. By
today, it is widely accepted, that majority of social phenomena
obeys non-linear properties or complexity. However, the research about
the origin of complexity in social sciences (including economics and
financial markets) has somewhat been unclear. 

The aim of this chapter is to show that economic policy and decision making is an application of the theory of complex systems. Hence, the economic systems can very seldom be reduced to the linear, forecastable systems, which is the essence of many economics and business textbooks. It seems that 
mankind has always searched for some \emph{clarity} or \emph{order}
in its arrangements. There has always been a drift towards
some hierarchy, or structure. Only in very recent past, perhaps due to the
progress of the so-called \emph{third industrial revolution} (cf. section \ref{sub:Trends-in-consumer}), there has
been some admittance that natural state of human society is better
described by complexity; non-Gaussian stochasticity or chaos. Or,
to put it in another words, the equilibrium economics has started to
leave its space to non-equilibrium economics.

This chapter is organized as follows: first the complex or non-equilibrium
phenomena is discussed in an economic arrangements from theoretical point of view. The rise of power-laws
is theoretically motivated. The change in the behaviour of consumers is discussed and the examples of power-laws are given. To conclude,
the guidelines are given for successful management of complex phenomena
in business and economics. Further, four applications based on non-equilibrium phenomena are discussed. The first application discusses
the opportunities for small economies in changing market environment.
The challenges of business managers is elaborated and the
chapter concludes with the two financial applications. The first of
them discusses threats arising from excessive usage of debt and the
second highlights the importance of risk management in the future.

\section{Complex social systems}
\label{sec:complexSSystems}
% Always give a unique label
% and use \ref{<label>} for cross-references
% and \cite{<label>} for bibliographic references
% use \sectionmark{}
% to alter or adjust the section heading in the running head
The population has rapidly grown in past centuries. So
have the social relationships between humans beings. They form increasingly
complex networks that have been studied by many fields of science
(e.g. complexity, network, chaos science). As it emerges, many, if
not all, human networks are scale free, i.e. they cannot be characterized
by some average or variance. For example: the question about average
company does not make sense, as in every industry there are giants
and pygmies side by side. Or perhaps the better example is the distribution
of wealth: already according to Pareto (cf. \cite{pareto}), the 80\%
of the land is owned by 20\% of the people. The spreading of such
systems is striking: internet routers, airport densities, book sales,
brand awareness are just few examples that obey so-called \emph{power-law},
i.e. system members obeying property $x$ decreases with $x$ in power
of $\alpha$. Mathematically the power-law can be written: 

\begin{equation}
P(x)\propto x^{-\alpha}
\end{equation}
where $P(x)$ denotes cumulative probability distribution of variable
$x$. The power-law leads to the \emph{scale invariance} that can
be explained as the system without characteristic scale to measure
it. 

It is important to note that the complexity can be driven from ordinary
functional dynamic equations that just contain the non-linear element.
Therefore the researcher, economist or businessman needs to be very
careful when building his/her theory or business plan on some equilibrium
in the system. Non-linear dynamics typically leads to non-equilibrium
systems; or to the equilibriums that are not stable. The important
question is to ask why? Why social systems lead to the power-laws,
scale invariance and chaos? In the following sections the complexity is theoretically motivated and the applications are offered for economic decision and policiy making.

\subsection{Motivation of complexity in social sciences}
\label{subsec:2}

It has been said that a biological system is in equilibrium only when
the cell is dead. The question arises, what determines, if an economic
system is under equilibrium or not? In the Table \ref{tab:Sources-of-non-equilibrium},
the conditions for certain equilibrium are compared with equivalents
of non-equilibrium. 

\begin{table}
\caption{\label{tab:Sources-of-non-equilibrium}Sources of non-equilibrium
in socio-economic processes}

\begin{tabular}{|p{3.5cm}|p{3.5cm}|p{4cm}|}
\hline 
Equilibrium & Non-equilibrium & Complexity resulting from non-equilibrium\tabularnewline
\hline 
\hline 
Deterministic linear dynamic equations & Deterministic non-linear dynamic equations  & Deterministic chaos\tabularnewline
\hline 
Converging variance of random process & Diverging variance of random process & Power-law distribution\tabularnewline
\hline 
Independent random variables & Time-related random variables & Fractality and long-term correlations\tabularnewline
\hline 
\end{tabular}
\end{table}

The complexity can be easily increased by combining any of the factors above.
For example, a stochastic process with power-law distribution and
long-range memory is called multi-fractal process (cf. \cite{mandel82,mandel97}).
As noted above, there is a number of reports about economic systems
obeying complex dynamics. However in economics, one should always
ask why the system is having non-linear properties? The key difference
between social and natural sciences is that the latter has super-universal
properties (laws of nature) that are present at any time in any place.
The social systems, according to the the best available knowledge
do not obey such universal properties. Even if such universal
law exists, then it is so vaguely defined, that the dynamical equations
cannot be derived. For example, the \emph{utility theory} suggests
that all economic agents are striving for maximum utility. The utility,
however, is not universally defined. Some consider money as a proxy
for utility, some add \emph{soft} factors such as happiness. Poincar\'{e}
has suggested that if the assumptions of mathematical models are not
valid, the models do not work. The same applies to applications
in economy and social science. Barab\'{a}si and Albert have shown (cf.
below), that power-laws are arising from growth and preferential attachment
of the system. Therefore, we can conclude, that the power-laws are
justified in the studies of \emph{current} economic conditions.
But due to the lack of super-universal social laws, one should always
be critical of applying the model for future forecasting purposes.
Additionally, on has to evaluate the assumptions behind each
model. If a socio-economic model is empirically verified from past
data; and it is claimed that this model is universal (i.e. applies
also to the future or is good for making predictions) then it can
be classified as \emph{historicism} as elaborated by Karl Popper (cf.
\cite{PopperHist}). 

To motivate existence of power-laws in social sciences one has to
go back into 1950s when Hungarian mathematicians Paul Erd\"os and Alfred
Renyi studied first the random graphs (cf. \cite{ErRe}). Their work
has guided studies of complex networks for decades until increased
research in the field started to question whether it is right to assume
complete randomness in real networks (such as internet) or expect
some organization of the system. If the network was random then the
distribution of the number of connections from arbitrary point in
the network (denoted as degree distribution) would follow Poisson
distribution. In 1999 Barab\'{a}si and Albert reported (cf \cite{barabasi}
and references in \cite{albert}) that (degree distribution of) real
networks deviated significantly from expected Poisson distribution
and obeyed a power law instead. They also showed that the system must
have \emph{growth} and \emph{preferential attachment} properties to
result power-law behaviour. Preferential attachment denotes that the
relationships between system members do not appear randomly, but new
links prefer to attach to connected members.

The complexity arises from non-linear components in the system. The
sources of non-linearity (but still deterministic and non-stochastic
behaviour) are countless: from micro-economics the supply and demand
functions can be non-linear (cf \cite{bischi}); production cost does
not have to be linear function of volumes (economists are calling
this \emph{scale-effect }or \emph{scale-efficiency}); and relationship
between unemployment and inflation (so-called \emph{Phillips curve})
might be non-linear to mention just the few. \emph{The power-law behaviour
is the signature of underlying complexity of the system}. It is the
key warning signal, that the system under consideration might have
some intrinsic non-equilibrium properties. 

The discussion above has shown, that socio-economic system can easily
become complex system. The presence of power-laws, memory and deterministic
chaos in economic process has been motivated and empirically confirmed
by many authors. Next it is discussed whether the global
economy has recently increased its complexity and therefore the scientists
and economists and businessmen have to change their paradigms in interaction
with surrounding economic problems. 

%\begin{quotation}
%Please do not use quotation marks when quoting texts! Simply use the \verb|quotation| environment -- it will automatically render Springer's preferred layout.
%\end{quotation}
%\quote{tsiteering}

\subsection{Trends in consumer behaviour}
\label{sub:Trends-in-consumer}

The special editorial report devoted to \emph{the third industrial revolution} was published in the April 21st, 2012 issue of the British weekly \emph{the Economist}.
The Economist claimed, that current changes in global economic development
have exceeded the usual evolutionary process and the global economy
might be in the verge of the revolutionary changes. This is, of course,
a revolution in production methods, not a political revolt against
ruling classes. 

The first so-called industrial revolution took place about two hundred
years ago, in the late 18th century in Britain with mechanisation
of textile industry. In the following decades the idea of letting machines
do the work (instead of physical work by people) spread around to the
other industries and countries. The second industrial revolution started
at the beginning of the 20th century in America and was characterized
by mass production. The iconic industrialist Henry Ford has said that
\emph{any customer can have a car painted any colour that he wants so long as it is black} (cf. \cite{henryford}).
In other words, the production processes were improved
and tremendous efficiency gains were achieved over the 20th century.
More people than ever before could allow themselves a car, computer,
washing machine or any other industrial product. Since the beginning
of the industrial revolution, the improvements in quality of life
for ordinary people are incomparable with the developments known since
the beginning of the civil society. 

By the beginning of the 21st century the global economy has reached
the point where production is no longer an issue. The unit cost of
production (in terms of raw material, time or labour) is lower than
ever before. But the problem has turned and the production issues
have been replaced with marketing issues. The supply of goods is saturated
and the consumers are more selective than ever before. Consumers
want to have individual solutions for the price of mass production
and all this should be available instantly. Auto-mobile industry,
the flagship of the second industrial revolution, does not produce
for the retail consumers identical cars any more. The keywords for
the third industrial revolution are: customer centric, flexibility,
speed, preciseness. The consumers are now literally dictating the
market. The importance of intermediaries and whole-sellers (who depict
the choice for consumers) is decreasing in time. Thanks to the possibilities
of e-commerce, even the physical constraints do not matter any more.
Everyone can shop online. 

These trends have major (or revolutionary) effect to the production.
The efficient manufacturing of big quantities is not suitable for
changed consumer behaviour. Not only the production quantities, but
the logistics should match the changed environment. \emph{The Economist}
makes case that the productions is therefore returning to the proximity
of consumers. Consumer preferences are subject to complex decision-making.
Therefore, the demand side of the economic supply/demand relationship
is expected to create clustering (i.e. power-laws). In addition, the consumer choices are very sensitive to the small details. In physics,
it denotes the sensitivity to the initial conditions, that leads to
the chaotic behaviour of the system. 

\subsection{Rise of power laws in the World economics and trade}
\label{rise_of_powerlaws}

Further it is analysed how recent societal trends have increased the complexity. The model of Barab\'{a}si-Albert (cf. \cite{barabasi}) is used. To recall, the model claims that the system must have properties of \emph{growth} and \emph{preferential attachment} to result the power-law behaviour. It will be showed that many social and economic systems obey those two properties and therefore the power-law behaviour is theoretically motivated. 

First the growth component is discussed. In order to motivate the complexity, the system must have the change in its size. To put it inother words, the number of social agents must vary in order to have power-law in social system. It will be quite easy to show that this is the case in many of the social systems. The first and foremost, the population of the Globe has reached 7 billion by 2012; and the
rate of increase as been also increasing. From a pessimistic note from economic perspective,
the increase has not been in the most developed parts of the Globe
and therefore it might not contribute to the increased complexity. This note can be ignored because of the decreased barriers in the global
trade. The World is becoming increasingly global in its trading affairs.
The local differences are fading away and therefore the increase in
population is reaching to the global markets. Therefore, the increased
trading relations also increase the number of potential customer for
each company regardless of its physical location. As it will be shown
later, this brings unique opportunities for the smaller countries. 

The second requirement from Barab\'{a}si-Albert model for arising the
power-laws is \emph{preferential attachments}. This phenomenon denotes the
behaviour, where social agents prefer to connect with the agents that
have more connections at first place. When intuitively true, it can
be proved by ruling out the opposite. Let us consider an arbitrary
social system, that is related to the human behaviour. Preferential
attachment does not exist, when the decision making over the population
is completely random. Whereas there is definitely an element of randomness
involved, any of the product or service for the sale contains an unique
value proposition that influences decision making. The examples of
those are: historical habits, expensive marketing campaign, cheap
price of the product, well-known (and prestigious) brand, lack of
alternatives and others. With the growth and preferential attachment components in place, the Barab\'{a}si-Albert model has its pre-conditions in place and the systems under observation obey power-law. This is a matter of fundamental importance in social and economic affairs: with the rise of power-law the systems become scale invariant, i.e. they loose the meaning of properties of average and standard deviation. 

The power-law in stock price fluctuation is caused because investors
have their preferences to buy stocks by definition (hopefully nobody
picks stocks randomly) and the number of investors in given company
is continuously changing (growing or decreasing). The number of internet
websites and academic papers grows; and is is likely that more popular
websites or papers are getting more and more connections to them.
Number of words in any language is growing; people prefer to use only
limited amount of them and therefore the distribution of word usage
follows power-law (i.e. Zipf's law). The wealth of the individuals
grow; but more money in absolute terms is earned by the ones with
higher initial capital (although the percentage of the return might
be higher with those of low initial capital). The list can easily
be prolonged and the power-law is observed in very many social systems.
But before the usage of its applications, it would be still wise to
consider, if the system has the instrinsic properties to yield power-law
behaviour. 

As a final note, the preferential attachment refers to the free will
of the agent. This is of course the cornerstone of free will (in political
terms) or free capitalism (in economic terms). The changes in political
regimes in the past decades have positively contributed to share of
people who are currently part of global capitalist system. Those people
have started to express their free political and economic will. And
the latter has significantly contributed to the rise of the demands
from consumer side. Coupled with the technological advancements, inter-connectivity
due to the internet, the global preferential attachments have most
likely gone through qualitative change that gives the rise of complexity
and power-laws in countless social applications. 

\subsection{About social predictions and economic forecasts}
Forecasting and socio-economic determinism has been a desire for the mankind since Plato. Karl Popper, in his seminal book of \emph{The Open Society and Its Enemies} (cf. \cite{Popper_OpenSoc}), analyses brilliantly the problems arising from unvertainty and openness of the society. According to the Popper, Plato saw and understood very well the changes in the society that have started to transform from tribalism to the democracy. Plato's response was to freeze all of the changes and return to the old, closed society. The key difference between \emph{open} and \emph{closed} society is that in the former the individual decisions of the people almost did not exist; all of their choices were pre-determinied by customs of the tribe or society. But this was also mental relief, as people did not bear any responsibility for such decisions. In open society, on contrary, people have to decide by theirselves about their actions, but they are also responsible for their actions. During the course of past two millenia, the problem of Plato has re-occurred a number of times (cf. numerous references in Popper's \emph{The Open Society and Its Enemies} \cite{Popper_OpenSoc}) with the issues circling around $(i)$ individual freedom and responsibility of the person and/or $(ii)$ collective welfare of the nation and the means to achieve this. 

It can be discussed, wheather the tribalism was truly closed society or not (at the end of the day, the laws of nature still applied to the society), but it has no relevance in the current context. The one and truly relevant point is that the society is a complex system with countless interconnections and non-linear relationships. It cannot be reduced to the linear system for the forecasting purposes. The power-laws and scale invariance (i.e. lack of characteristic measures such as averages or standard deviations) simply cannot allow to ignore one-off, big time changes. \emph{Critical Mass}, a book by Philip Ball (cf. \cite{ball}), describes wide variety of such phenomena. \emph{Fooled by Randomness} and \emph{Black Swan} by Nassim Nicholas Taleb (cf. \cite{taleb_fooled,taleb_swan}) adds the flavour of the financial markets. 

\subsubsection*{The problem of the Analyst}
Many fields of human activities celebrate calendar year by nominating and choosing the best performer of the year. Among others, the best athletes, artists and architects get selected. From another perspective, various magicians are continuously providing the \emph{forecasts} for the coming periods. Regardless of the used methods (tarot cards, celestial bodies or other \emph{tools}), one can be very sceptical about the social forecasts, as once again, there are no super-universal laws that describe the social systems. However, all kinds of forecasts are very popular, as they provide a sort of security that people would like to have in their lives. This is especially true for the stock markets, but also for various aspects of people's personali lives. A very interesting book about forecasting  (cf. \cite{rowland}) was written by Ian Rowland, a British magician, who has debunked various aspects of paranormal phenomena. 

What is the role of forecasting in economic decision making? Certainly the stock exchange prices are driven by various forecasts: the ones drawn by the companies themselves and the others offered by analysts in investment banks. In addition, the micro economic (or company level) forecasts are dependant on the general macro economic situation; and further the predictions of economists influence the forecasts of the stock analysts. Note, that the current passage is by no means restricted to the stock exchange, but the logic offered can be easily applied to the other fields of finance (e.g. agriculture as discussed in section \ref{sec:Application:-Negative-impacts}). But are such predictions contributing to the economic decision making, or perhaps \emph{vice versa}, make things worse? 

As discussed in section \ref{subsec:2}, the mathematical models should be dropped, if the assumptions are no longer valid. Further, the only super-universal law in economics can be formulated as follows: 
\begin{equation}
Revenues - Costs = Profits
\end{equation}
Therefore, the only reasonable economic forecast relies in simple modelling, where revenues and costs are depicted. If the model works on paper, it may also work in real life. If the model does not work on paper, then there are very little chances that the business will work in real life. However, typically economic models contain non-linear inputs because of stochasticity (variability of input and/or output quantities), non-linearity (clustering of purchases or sales of the company) or other reasons, that make the forecasting pointless. For example, in order to forecast stock price the analysts must account not only for the company-specific reasons, but as well as behaviour of other investors in the market. This equals of forecasting the social (i.e. complex) systems. 

The problem of the analysis broadcasted in popular media includes the following shortcomingis: $(i)$ missing error estimates, $(ii)$ missing back-tests of the achieved results; and $(iii)$ missing personal responsibilites. In physics, all experiments are conducted with mandatory error estimates. In economics and financial markets, the error estimates are never given that would leave the user of such forecast with no idea of the accuracy of the forecast. Further, even if the forecast proved to be right, there is never no information given about the method itself, or the reliability of the forecast in slightly different initial conditions. From section \ref{subsec:2}, the complex systems are very sensitive to the initial conditions. Therefore, if no additional information is given, there result is not qualitatively different from the random luck. Thirdly, no analyst is responsible for their results, that are used in media. So, even if the results are not correct, the analyst typically bears no responsibility for the advice. Why should anyone take such forecasts seriously? 

The critique above was not intended to dismiss all aspects of economic analysis. As noted previously, the models that build on the pre-determined rules can be verified and falsified (as defined by Popper in \emph{The Logic of Scientific Discovery} \cite{Popper_LSciDisc}). However, the critique is addressed to the blind belief of any of the forecasts of complex social systems.

\subsection{Possibilities of handling the complexity}

The power-laws and chaotic processes have already put many companies
out of business and most probably will do also in the future. In this
section some stylized approaches are suggested in order to manage
the changed economic environment. 

\subsubsection*{Openness and international competition}

The most fascinating thing about the globalization and international
integration is the international competition. The competition, by
definitions means threats and opportunities. The competition forces producers to continuous improvements in order to appeal the consumers. The definition of the consumer and the market has changed
in the past decade; so has the demand of the consumers. About 20 years
ago, the consumer potential of any product was determined by population
living in certain area and having relevant income level. Due to the advances of internet, the geographic
borders have vanished. The physical stores are there to remain for
food and other daily products (and perhaps daily services such as hair-dressers, car washes etc). On the same time, the dealerships for consumer staples
have to reconsider their business plan. The market for all
products and services has become more competitive and therefore the following
can be generalized: the producers must aim to produce globally best
product since the consumers want to buy globally best products. Obviously,
the definition of the \emph{product} can include various components;
as described by value proposition in (cf. \cite{BlueOcean}). If
the local language is the part of the value proposition (for example
for a video game), then the producer can claim, that he/she is doing
the best product in local language; that sells with higher price than
the English language equivalents. This might be true, and the producer
might be successful, but in conventional market conditions, the producer
always bears the risk that the consumer's preferences change and s/he
is out of business. Therefore, given the rise of competition and higher
demands from consumers, the producers must always be open to the innovation
and assume the global competition; even if they are servicing only
local consumers. 

\subsubsection*{Tolerance and variety of ideas}

Hayek has said (cf. \cite{hayek}) that the society cannot be measured
in a single scale of more and less. The number of opinions might be
as large as the number of people. The democratic society respects
all opinions that are not dangerous to others. The dominance
of the single opinion is also dangerous as it can be wrong. In the
context of economic management, due to the complexity, the direct
and indirect outcomes of decisions are never clear. Therefore it
is important to consider and tolerate within society or company, but
also in department or family level the variety of opinions in decision-making. The next question arises from the implementation
of the decision. Should it be done with one \emph{big bang} or slowly,
step by step. As it was discussed above, the complex systems tend
to be non-equilibrium. So, every big step may drive the unit under
discussion quickly out of equilibrium. Therefore it is important to
manage step by step. In popular reading, it is also known as method
of trial-and-error. Should the idea or decision prove to be wrong;
one can easily reverse the situation and try other solutions. 

The pessimist in complexity phenomena might argue that the piecemeal
implementation of any plan never allows to achieve extraordinary results.
A wise person never puts all eggs into one basket. Whether the
company is an established one or a start-up, the owners very seldom
take single risks with all of their capital. If the new business venture
is still pursued, with all the capital under risk; the successful
outcome cannot still be classified as wise, but rather lucky. 

As the final note, the variety of ideas very seldom rises from single
source of knowledge. Therefore, along with continuous innovation,
the variety of ideas should be always searched for. One original sources
of the ideas is to look at the intersections of various disciplines.
People with different background are more likely to produce truly
innovative ideas as opposed to similar people. Interdisciplinary fields
of science, business or social life are more likely to yield new ideas
that expand mankind's knowledge and welfare.

\subsubsection*{Low dependence on external help}

External interference into any physical system disturbs the system.
As discussed previously, the small disturbances can yield the qualitatively
different outcome of the system under chaos. But what if the disturbances
are permanent and/or large? Then the system has to be redesigned and
all of the dynamic equations should be rewritten. The social systems
behave analogically. The small disturbances are perhaps new competitors;
or small new technological innovations. The technological innovations
can lead also to the new industries and big changes, but this is also
part of the usual market behaviour. However, the government interference
into economic system is big disturbance that reshapes the whole economy
in general or any industry in particular. The government interference
creates market distortions that companies should comply with.
This can be advantageous (in case of subsidies) or disadvantageous
(rules and regulations). Important is to note that companies under
global competition may respectively have opportunity or threat to
survive. The government interference is very powerful tool that influences
the system. Under complexity, it can therefore with the fraction of
the second, reshape the destinies of many companies and workers in
those companies. The same is true also for the demand side of economy.
With the increased regulations the consumers might pay higher prices. 

To summarize, the companies should be aware of the distorted market
models, since they are competing internationally. This is
also the case, if the distortion is beneficial in short term, and
is creating the advantages. The threat is that the managers might
be deluded from observing international competition; and if the
public support disappears, they may find themselves in trouble.

\subsubsection*{Qualitative improvements in risk management}

The increased complexity demands the qualitative improvements in risk
management. There are various sources where the company or private
individual can be exposed to the complexity. It is utmost critical
that the company will be aware of such phenomena. It has to understand
the sensitivity of its cash-flows of the variables it is exposed.
Further, the risk mitigation plan should be devised. It is important
to stress, that not all risks can be or should be hedged. After all,
the companies typically earn money for risk taking. Hence, the good
risk management is not risk minimisation, but risk optimisation. But
one can optimise only after the risks are identified and quantified.
This requires dedicated attention from to top management or owners
of the company; after all, it is their interest that the company would
not suffer against \emph{unforeseen} risks. And all of the risks arising
from complexity are not seen without proper attention. The risk management
issues are also discussed in section \ref{sec:Application:-Negative-impacts} of this chapter.

\section{Application: Opportunities for small economies}

Countries, like companies are obeying power-law, if their size
(in terms of population or economic output) is observed. It is tempting
to ask that who will win from the \emph{third industrial revolution}
(cf. section \ref{sub:Trends-in-consumer}) or from complexity or
combination of both. Instead it is asked, what are the opportunities
for small economies to benefit from the underlying changes in
the global economy. Note, that the definitions of \emph{economy} and
\emph{small} are not given in this context. It can apply for the nation,
company, or family. 

The value proposition of the production company in the new environment
can be elaborated as follows: first and foremost, the changes in consumer
behaviour call for \emph{individual solutions}. In every industry
the consumers are expecting the tailor-made solutions that are different
(or at least look different) than the others. This calls also for
\emph{decreased} production \emph{quantities} which, in turn, affect
production processes. Therefore, the \emph{flexibility of production}
is playing increasing role in global competition. The producer or
service provider has to be quickly able to adjust to the new orders;
the delays in production drive up production costs. The flexibility
has also other meanings: it is ability to quickly introduce new products
or variations of existing ones; it implies quick and \emph{reliable
delivery} of the goods; it raises a question of whether the company should
enter or exit new elements in product value chain. This means that
the company may decide to start to produce more value-adding components
to existing products; or to increase the production cycle (i.e. start
producing also goods that it was previously buying in). Excelling
in production does not necessarily make the company successful as
the products also need to be sold. But selling has not that much changed
due to the complexity. There is just another dimension; that is to
provide the value proposition above: flexibility, individual solutions,
small quantities, efficient production and quick delivery. However,
since the competition in production side is also growing the reliability
and \emph{business ethics} of each company as well as nation starts
to play increasingly large role. The successful companies cannot allow
themselves to break their promises or violate oral or written agreements.

The concept of flexibility has so far only limited reach in academic
literature. Only very recently, a group lead by prof. Luciano Pietronero
has devised a new method to rank the countries. It opposes two-hundred-year
old concept (cf. for example \cite{marginalism}) of economic specialization (and of static equilibrium)
and ranks the countries by complexity of their products (cf. \cite{pietronero}).
The suggestion of their method is to study the complexity (and variety)
of the exported products that serves the proxy to the future well-being
of that country (as measured by GDP). 

The small economies can use this change in economic landscape for
their benefit. It is clear advantage as compared to the big economies,
that are used to make huge quantities of similar products. Obviously, bigger economies adjust as well; but their production efficiency
advantage is disappearing. The small economies have to be agile in
finding new opportunities. They have to strive for the best product
in the world and not for a less. Continuous innovation and rethinking
of the business model is of benefit. They must continuously try
new things; if one does not try, one cannot also succeed. But the
risks have to be carefully measured. It is dangerous to get stuck
with single buyer of the production: not only this makes negative
impact to the agility, but also the buyer might start to push down
the margins of the producer. Finally, the small economies should be
counting only to themselves. It applies to the national level, but
also to the company level. No government subsidy will make any of
the companies to produce or sell better. From public point of view
it has to be assured, that none of the productions is discriminated
by other countries; and the government might want to help with investments.
But the companies have to find the customers by them selves. No government
support (by small country) will help to reshape the global consumer
behaviour. 

To conclude, the increased complexity (due to globalization, usage
of internet and liberalized markets) opens the opportunities for the
small economies. With flexible production, reliable business ethics
and managed risks, they have all chances to succeed. 

\section{Application: Changes in business management}
The fact that the world becomes more and more complex was already
discussed above. It is reasonable to assume that business environment
also changes and becomes more and more complex. 

\subsection{Problem of trustful sources: success-driven business literature ignores the failures}
There is a myriad of books written about business and management, that typically state
or study trivialities: buy cheap and sell expensive; in order to succeed
the unique value proposition needs to be developed and exploited.
There are plenty of tools to determine the strategy. For example, 
one only needs to draw Porter's \emph{five forces} (cf. \cite{porter}). In the course from determinism to chaos more
individual approaches start to prosper. Blue Ocean Strategy (cf. \cite{BlueOcean})
is the method where value proposition of individual product/service
is modified; similarly the methods depicted by Moskowitz and Gofman
(cf. \cite{BlueElephant}) about segmentation. However, what seems
to be very common for the conventional business books is absolute
ignorance towards survivorship bias: namely all of the examples brought
forward are positive. At least in theory (and much in practice) there
are failing business strategies. However, those seem to be missing
from popular literature: almost always the theories are backed with
the success-stories. To summarize, all of the (successful) books write
about successful strategies, leaving unanswered the empiricist's question:
what made a strategy fail? It would be interesting to speculate, how
many product launches, business ventures, restructurings and similar
has to fail in order to produce single successful one.

\subsection*{Unsuccessful response to the complexity}

In late 1990s, before the burst of so-called \emph{high-tech bubble},
the popular buzz-word was \emph{New Economy}. This term was coined
mostly to describe the technological advancement. But not only. It
was also symbol of boundary-less activity, agility, in some cases
loose financial management. For some time there existed a claim that
the company of the \emph{New Economy} does not have to be profitable;
and the right assessment of the value of the company (i.e. share price)
was not done by money earned to shareholders, but clicks received
by company's web-site (there was even price-to-click ratio introduced
to be a better proxy than price-to earnings ratio). By today such
party is clearly over; performance management is back in stage and
companies are counting money more than ever before. As an aftermath
of recent economic and financial crisis, the regulations and standards
in many sectors (including financial, industrial and services) are
about to increase. The companies have to strive for order in management
and market relationship in order to survive. 

Yet, the 21st century has not made the shareholders rich. The performance
of the biggest stock indices in US and Europe has been modest, if
not negative, in the past decade. If not nominally, then volatility-adjusted
performance for sure. (Note that the aim of this chapter is not to discuss
the statistical properties of stock indexes; there are most definitely
market segments that have posted life-high returns). At the same time companies around the globe have succeeded in many paradigm-changing
innovations: the usage of Facebook, smart-phones, hybrid cars or renewable
energy has increased by order of magnitude. It would be probably hard
to find a single corporate executive who can claim that their efficiency
has not risen. Additionally, the financial management of companies
(and public sector) is better than ever before. The market has done
its job perfectly: companies are fitter than ever before. But as discussed
above, the consumers are also demanding more than ever before. The
complexity has risen, but the response has not yet been successful.

\subsection{Through Chaos to Determinism}
\label{sec:chaos}

To summarize at this point: corporate executives (as well as public
administrators) have to respond to both: $(i)$ increased consumer demands
for value innovation (that drives up costs); and $(ii)$ increased demands
from regulatory and financial institutions. And, as companies become
more and more competitive, there is the third dimension: ongoing competition
for the labour force. What is the right management approach in this
context? Tom Peters describes in his classic book In Search for Excellence
(cf. \cite{Peters}) various management styles that make companies
successful. However, it seems that all of his advice is bound to linear
evolution - that should be amended in the context of non-linear or
non-equilibrium economics.

\subsection*{An illustrative model of chaos in economics}
In this section an illustrative model is constructed in order to demonstrate the emergence of chaos from simple micro economic relationship. 

Consider an arbitrary company that sells an arbitrary good. It is simplified that the company is selling the unit labour hours; and the quantity available for sales (i.e. supply) at time $t$ is denoted as $q_S$. In this model, the quantity for sales can be also interpreted as the satisfaction of the employees - the bigger the number of employees the higher is the satisfaction of the employees. The customers are purchasing the same unit labour hours and their demand (at time $t$) is denoted as $q_D$. It can also be interpreted as the satisfaction of the consumers - the more they are buying the higher the satisfaction. At time $t$ the transactions are concluded at the price $p$; the wealth of the shareholders is $W$; and the employees are compensated at the unit rate of $C$. Therefore, the full cost of production is $q_SC$; the revenues of the company are $q_Dp$. From these simple definitions the following equations are derived.

The satisfaction of the employees is positively related to the compensation. The compensation (i.e. salary) has to be higher than critical value $C_0$; otherwise the employees are leaving. 

\begin{equation}
\label{eq:salary}
%\frac{dT}{dt}=\gamma(K-K_0)\dots
\frac{dq_S}{dt}=\gamma(C-C_0) 
\end{equation}

The satisfaction of the consumers is negatively related to the price. If the price is higher than critical value of $p_0$ the consumers will not conduct any transactions.

\begin{equation}
\label{eq_consumers}
%\frac{dR}{dt} = \delta(H_0-H)\dots
\frac{dq_D}{dt} = \delta(p_0-p)
\end{equation}

The relationship between shareholders and employees is also positive: if the shareholders are doing well, they tend to raise the salaries of the employees.

\begin{equation}
\label{eq:wealth-salary}
%\frac{dK}{dt} = \beta A\dots
\frac{dC}{dt}=\beta W
\end{equation}

If the supply is larger than the demand, the price has to be lowered. It is assumed that the price is not related to the wealth of shareholders.
\begin{equation}
\label{eq:price}
%\frac{dH}{dt}=-\alpha(T-R)\dots
\frac{dp}{dt}=-\alpha(q_S-q_D)
\end{equation}

Finally, the change in shareholders' wealth equals to revenues less costs. 
\begin{equation}
\label{eq:wealth}
%\frac{dA}{dt}=HR-KT\dots
\frac{dW}{dt}= pq_D-Cq_S
\end{equation}

Merging Equations (\ref{eq:wealth}) and (\ref{eq:wealth-salary}) yield following:

\begin{equation}
\label{eq:wealth2}
\frac{d^2C}{dt^2}= \beta(pq_D-Cq_S)
\end{equation}

To summarise, the micro economics of simple company can be presented in a form of system of ordinary differential equations.

\begin{equation}
\left\{ 
\begin{array}{l} 
\frac{dq_S}{dt}=\gamma(C-C_0) \\
\frac{dq_D}{dt} = \delta(p_0-p)\\
\frac{d^2C}{dt^2}= \beta(pq_D-Cq_S)\\
\frac{dp}{dt}=-\alpha(q_S-q_D)
\end{array}\right.
\end{equation}

\begin{figure}
\label{fig:RK}
\includegraphics[width=.49\linewidth]{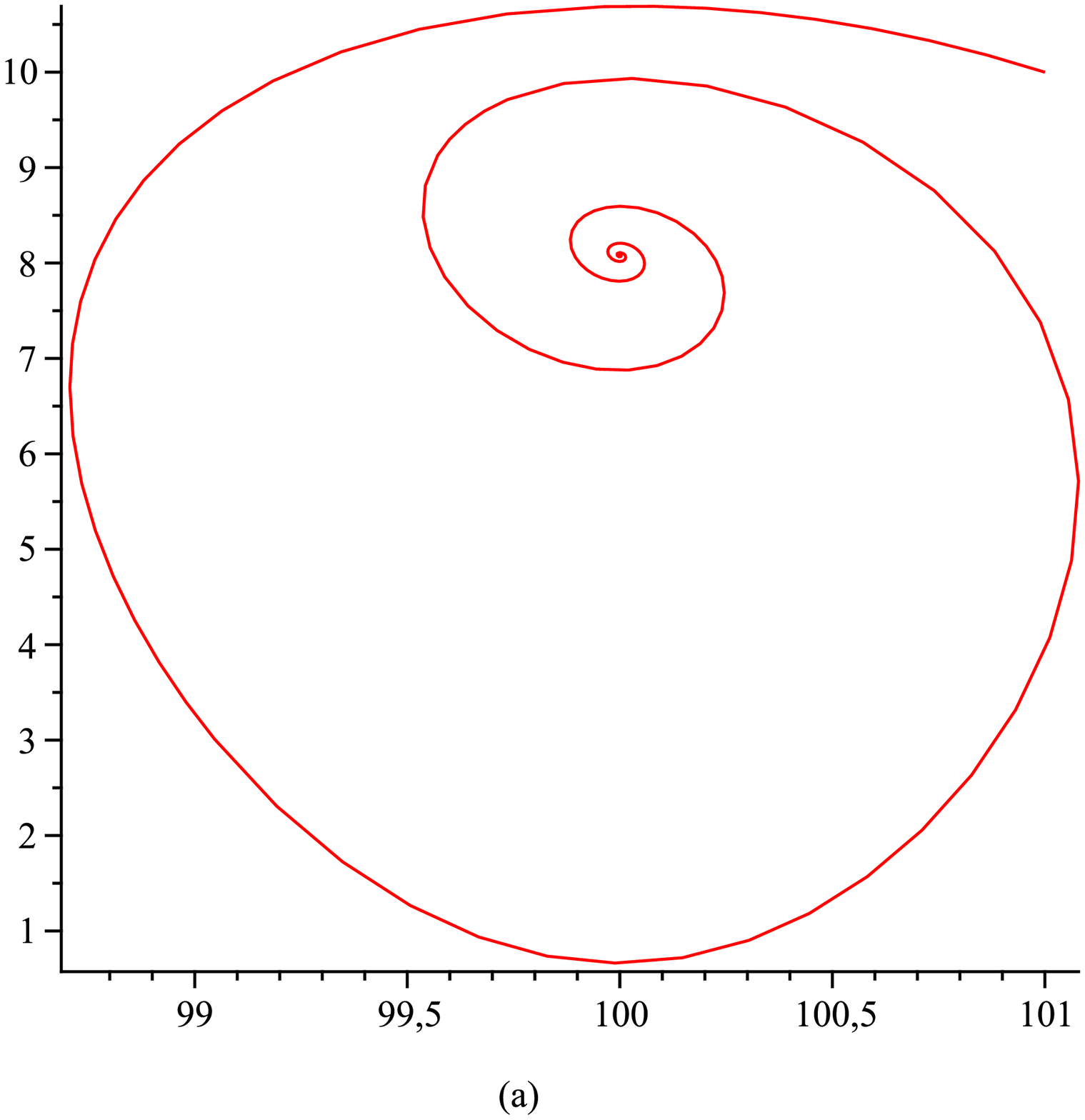}
\includegraphics[width=.49\linewidth]{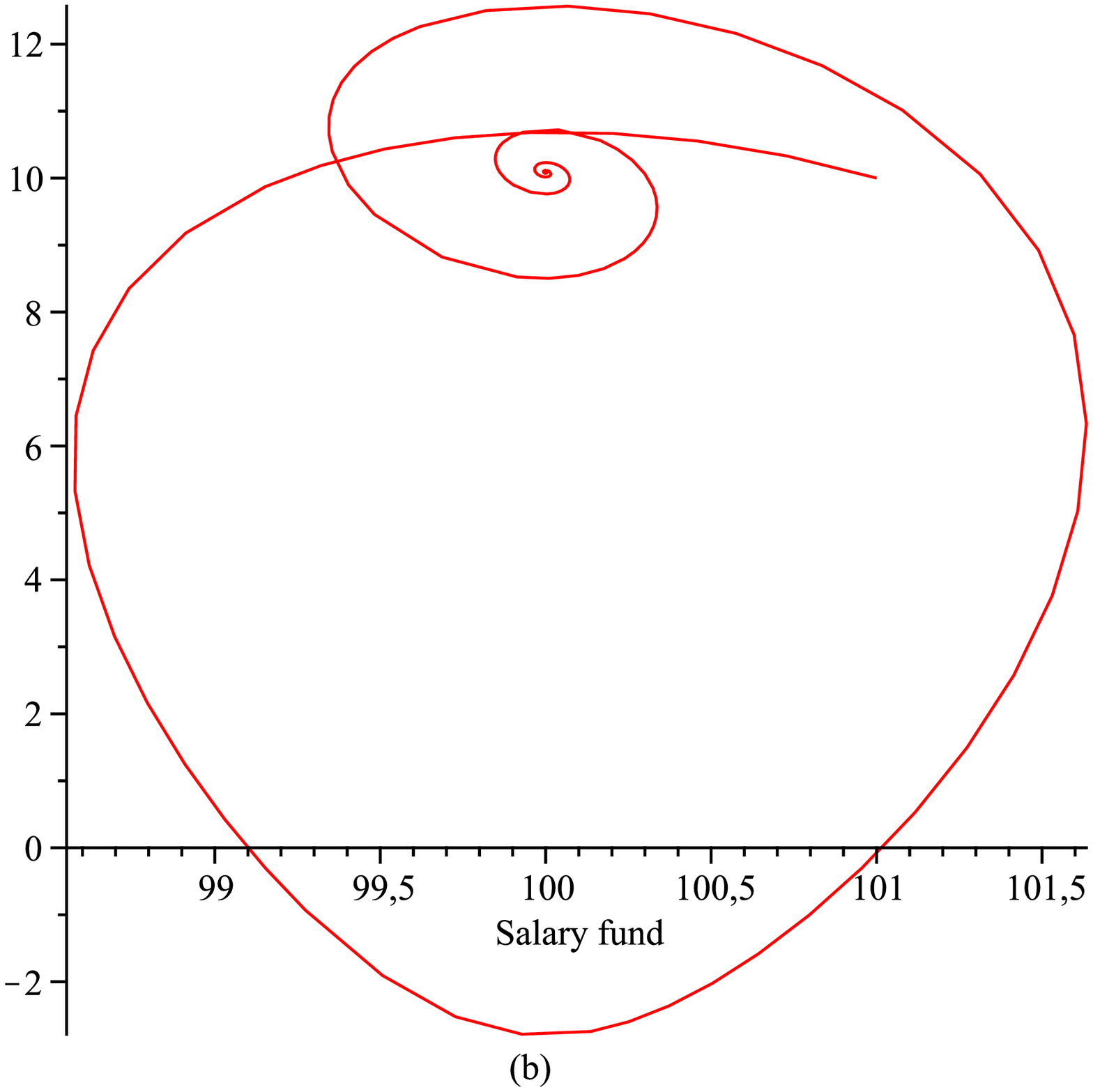}
\caption{$C$-$q_D$ plot with the parameters. $(a)$ $\alpha=0.6$; $\delta=0.2$; $\gamma=0.3$; $\beta=0.005$; $C_0=100$; $p_0=100$. Initial conditions of the system: $q_S(0)=10$; $C(0)=101$; $dC(0)/dt=-0.2$ and $p(0)=101$. $(b)$ $\alpha=0.525$; $\delta=0.2$; $\gamma=0.3$; $\beta=0.005$; $C_0=100$; $p_0=100$. Initial conditions of the system: $q_S(0)=10$; $C(0)=101$; $dC(0)/dt=-0.2$ and $p(0)=101$.}
\end{figure}

Due to the non-linearity, the emergence of chaos can be predicted. It is not attempted to solve these equations analytically. However, it is easy to demonstrate that the numerical solutions inevitably lead to the chaos. In figure \ref{fig:RK} the happyness of the consumers (variable $q_D$) is plotted against salaries $C$. From economic point of view the vertical axis demonstrate the quantities demanded by the customers ($q_D$). In both plots the quantities in demand drop to the low territories, but recover later. In chart $(b)$ the value of $q_D$ is even negative at some point. In real economic situations, this represents very bad business conditions and perhaps even the bancrupcy of the company. 

Although very primitive, the illustrations carried out in this section demonstrate very well the chaotic and complex nature of the business conditions. With smallest change of the parameter $\alpha$ the qualtitative difference between those two situations becomes apparent. Therefore the policy makers in social systems (including business and economics) have to be very careful with their decision-making. Only smallest change in any of the parameteres can lead to the qualitatively different outcomes. 

\subsection*{Chaotic management approach}
It is proposed to implement a \emph{chaotic management approach} in
order to succeed in the 21st century. Chaos, as defined in physics,
refers to the dynamical process that oscillates randomly around some
attractor (or basin). Chaos is not randomness. It is random only by
observing from short distance (e.g. short time). By borrowing the
phraseology from Graph theory (cf. \cite{albert}), the successful
company is consisting of graph of three nodes: clients (i.e. public),
shareholders and employees and three edges. This \emph{Corporate Graph} serves as the basin for attractor of chaotic dynamical process as
the company evolves during the time. The process gets very complicated
as there is a high degree of noise (news-flow, competitor and customer
behaviour) and high degree of conflict of interests across the edges.
As one can see, the determinism in the contemporary management is
unstable. Therefore the conventional tools of performance management
have to be adjusted. The goals and targets might change overnight
and the management culture has to account for this. The management under
chaotic conditions is much more complicated than the management under
deterministic or linear conditions. However, this is most likely irreversible
trend and the management in the future is most likely even harder
than the one of today. 

To conclude, the modern management is like deterministic chaos. Whereas
underlying structure of the stakeholders interest is determined (by
the Corporate Graph above) the daily management is full of randomness.
Successful leader has to keep its graph together with sound balance
between short and long time horizon. 

\section{Application: Threats from debt accumulation}

The essence of debt is to gear up the shareholder's
capital. The motivation and extent of debt usage may vary. Business
textbooks recommend to use it for capital optimization purposes, but
debt may also be required due to insufficient amount of equity in
the company. The sources of debt financing consist largely of two groups:
banks (typical for continental Europe) and debt capital markets (Anglo-Saxon
model). The same principles apply to the private individuals; only
the main reason for them is to engage into larger investments than
their cash accounts allows. Usage of debt is largely justified in
daily business processes and offering debt financing is major source
of revenues for commercial banks. Mismanagement of the debt leads
to the bankruptcy of private entity and/or credit losses for commercial
banks. Therefore, the debt issuing and management is of interest to
both: borrower and lender. 

In addition to private individuals and companies the States (via respective
governments) are issuing debt to manage public finances. Whereas there
are differences between the countries, the public debt tends to be
larger than the private aggregated debt. Again, in general the usage
of public debt is justified with the reasons being similar to the
private purposes. And similarly, mismanagement of public debt can
become a threat to the national solvency. 

From the textbooks, the main risks associated with debt are: credit
risk (i.e. borrower's financial inability or institutional unwillingness
to serve the debt), interest rate risk (i.e. difference between market
value and notional amount) and currency risk (i.e. difference arising
from currency fluctuations). 

\subsubsection*{Implications of complexity to debt management}

In must be noted, that by introducing complexity, there are no changes
in any of the matters above. Conventional private, corporate or public
finance has no alterations due to the complexity. What complexity
does, is making financial system more fragile. Namely, the usage of
debt in the balance sheet implies certain level of confidence that
borrower is able to serve its obligations. Under stable conditions,
the borrowers can estimate their cash flows and then derive their
debt service ability. Usually the cash flows of households
consist of salaries; the free cash flow of companies is revenues minus
costs; and the public sector cash flows is essentially collected taxes
plus one-off items, such as privatization. 

Complexity and power-laws add additional element of uncertainty into
cash flows. The economic system is never in equilibrium and therefore
the borrowers must qualitatively re-estimate their ability to service
the debt. All economic agents must ask the following questions: what if the salary/revenues
disappear from one day? What if there is unforeseen economic depression?
What if the assumptions of balance sheet optimization task are no
longer valid? The debt needs to be serviced at all given times. The
borrower faces problems, if it fails even in short term despite the
fact that the balance sheet and all other conventional measures are
still satisfactory. Complexity makes borrowing fragile since the short-term
problems can kill the borrower. As we have recently seen (cf international
help to Greece, Ireland and other countries in Europe) the sovereign
States are not that different from private entities. It can be speculated
that the increased amount of money has increased the complexity and that, in turn, has triggered the problems for over-borrowed
countries. 

To conclude, the usage of debt is very common and it helps to increase
the opportunity set for private and public institutions. However,
the extent of debt usage should be considered carefully; the
optimization techniques for certain aim (e.g. return of equity) are
subject to equilibrium that does not hold in real, complex economic
systems. 

\section{\label{sec:Application:-Negative-impacts}Application: Negative impacts
of market-driven complexity}

So far the discussion has mainly been focused on positive aspects
of complexity. This has opened many new alternatives and created new
possibilities for people. However, there is a clearly negative
application due to the market liberalization and the rise of complexity.
This is the application of long-term investments. 

The long-term investments such as manufacturing plants and infrastructure
have faced clear set-back from the liberalization of the markets.
Let us consider the simplified example of electrical plants. The life-span
of the plant and therefore the business plan is drawn for decades.
The feasibility analysis of the construction of the new plant goes
like follows: revenue minus cost minus debt service costs
yields the shareholder's profit. The output of the electrical
plant is by definition the electricity and hence the revenues of the
plant is generated from the sales of electricity. The cost of building
an electrical plant is huge; this can be done only by borrowing. From
the previous, borrowing makes the system fragile and the debt servicing
flows have to be carefully planned with buffers and cushions for
the extraordinary events. However, with the good investment plan it
is not unusual to draw the feasible business plan for electrical plant.
The problematic aspect is added by increased uncertainty in revenues.
Note, that the complexity arising from the market fluctuations of
input commodity (e.g. coil, gas) is ignored here because of simplicity.
By market liberalization, the electricity is freely traded in the
market. What trades in the market tends to fluctuate. And this adds
uncertainty and complexity by the order of magnitude. Since the amount
of equity in the balance sheet is limited; the drop in revenues can
influence the debt servicing ability. One should not be confused between
market revenues and revenues from public subsidies. Public subsidies
exist to eliminate some market inefficiencies; or to execute a
political goal. In the context of electricity such goal is to facilitate the renewable energy sources. Hence, there is stable stream of
cash-flows coming from government that is typically reliable source of revenues
in business plan. To conclude, there is probably a relationship between
market liberalization and willingness for private entrepreneurs to
set up new electrical plants. 

The example of electrical plants can be easily generalized to any
of the production facility. The business plan faces additional stochasticity
from almost all of the inputs as well as outputs. To elaborate: one
of the largest cost items is electricity, also various commodities
including metals and food. Note that the cost of labour is not accounted
here - the labour market has not essentially changed in the context
of recent jump to the higher level complexity. Similarly to the costs,
the revenue side (i.e. the sales of the company) fluctuates in the
market; and it should be as the output of one company is input to
other companies or individuals. Whereas the input prices can be passed
through to the output prices is the function of company management
skills and market practices. 

An interesting trend of increased market-based trading takes place in agriculture. Recently, the number of agricultural products that are
traded in the financial market has increased. This has lead to 
new market participants in the form of financial investors (growth in
map). From consumer demand, there is also the preferential attachment. Recall from section \ref{rise_of_powerlaws} that growth and preferential attachment are the two pre-conditions for the rise of power-laws from Barab\'{a}si-Albert model.
Therefore, the rise of power-law has gained its legitimacy and, indeed,
the prices of agricultural commodities have started to fluctuate in
more unpredictable way. Note, that the prices used to be predictable
as the market participants included only producers on supply side
and consumers on demand side of the economic equation. Therefore,
the complexity of agricultural business model has exceeded the production-specific
aspects and includes now also the complexity of the financial markets. 

What would be the tools for businesses to overcome the complexity
of markets? The first and foremost, the business managers have
to accept the complexity and to admit the higher degree of unpredictability
in their business plans. The second line of defence is to explicitly
state the quantities under risk and then to find the risk mitigating
solutions. For example, the electrical plant managers should think
about their cash-flow sensitivity towards the price of gas or coal
(inputs) and electricity (output). The risk mitigation can be done through
financial contracts as OTC forwards, futures; options or other similar instruments;
or by passing through the input fluctuation to the output. The problem
with this approach is that the length of such contracts is typically
short and cannot be used for the whole lifetime of the business plan.
Despite of that, the increased complexity calls for the qualitative
increase in risk management of the corporate and/or public management.
Another solution to the problem is financial innovation. Namely,
the loans of the banks might depend on the price of some commodity,
but it is only partial solution as then the banks would be taking
the price risk. The full solution is that the banks can also finance
themselves based on some commodity index; that in turn calls for the
financial market to create such instruments. And, as we have seen, this might
trigger in turn an additional level of complexity.

\section*{Conclusion}

Non-equilibrium economy via its applications of chaotic dynamic and
non-linear stochasticity is in the rise. The rise of power-laws in
many social systems are discussed and it is shown that the systems with growth and preferential attachments are characterised by power-laws. 
These conditions are satisfied in increasing
number of fields in socio-economic landscape and therefore the non-linear
or complex phenomena is increasingly dominant in social systems. In
addition, the changes are also under-way in global consumption patterns
that together with inter-connectivity through the internet make business
and economic environment more and more complex. For successful management
under complexity, a following principles are offered: openness and
international competition, tolerance and variety of ideas, self-reliability
and low dependence on external help. 

Despite of increasing complexity, it seems that small economies have
good prospects to gain from the global processes underway. The key
to success is flexible production, reliable business ethics and good
risk management. Management itself is also changing. The static approaches
and tools have to be complemented with dynamic and more agile approaches.
Corporate executives do not have the luxury of not to react to the market
information promptly. The managers have to find good balance between main
stakeholders (customers, shareholders and employees) as well as market
reactions. From financial aspects the excessive
usage of debt is questioned. The debt makes companies fragile as short-term temporary
downturns can under unfavourable circumstances kill the company. The
increasing non-linearity in the economic surroundings has influenced
many industries. As it is shown, the business models start
including also the financial risk management as an integral part of
company's operations.

\begin{acknowledgement}
This contribution was written in the 2nd Ph.D. School of ''Mathematical
modelling of complex systems'' in Pescara, Italy in July 2012. Author
would like to thank all of the attending students and professors for
fruitful discussion; and especially Proffessors. G.I. Bischi and T. Bountis. Author would
also like to thank Prof. J. Engelbrecht and Dr J. Kalda from 
Institute of Cybernetics at Tallinn University of Technology for fruitful
discussions. The support of Estonian Science Foundation (Grant ETF7909) supported by the EU through the European Regional Development Fund is highly appreciated. 
\end{acknowledgement}

%%%%%%%%%%%%%%%%%%%%%%%% referenc.tex %%%%%%%%%%%%%%%%%%%%%%%%%%%%%%
% Robert Kitt. References of the "Application of Complex Systems in Business and Economics"
% Chapter submitted to Springer book Chaos Theory in Politics
% (C) Robert Kitt, December 25th, 2012

\end{document}